\documentclass[12pt,a4paper] {article}
\usepackage[cp1251]{inputenc}
\usepackage[T2A]{fontenc}
\usepackage[english]{babel}
\usepackage[dvips]{graphicx}
\usepackage{amsmath}
\usepackage{amssymb}

\usepackage{geometry}
\newcommand{\ds}{\displaystyle}
\def\bR{\mathbf{R}}

\begin{document}

\begin{center}

{\bf \large Regions of Possible Motion in Mechanical Systems\footnote{
Printed in USA. Soviet Physics Doklady, 1982, vol. 27, pp. 921--923.\\ Translated from
Kharlamov M.P. ``K issledovaniju oblastej vozmozhnosti dvizhenija v mehanicheskikh sistemakh'', Doklady AN SSSR, 1982, {\bf 267}, 3, pp. 571--573.}}

\vspace{2mm}
{\bf M. P. Kharlamov\footnote{Donetsk State University (1982).}}
\vspace{3mm}

Presented by Academician P.Ya.Kochina March 5, 1982 \\
Received April 4, 1982

\vspace{3mm}

\href{http://adsabs.harvard.edu/abs/1982SPhD...27..921K}{http://adsabs.harvard.edu/abs/1982SPhD...27..921K}

\end{center}

\vspace{3mm}

We consider a dynamical system
\begin{equation}\label{eq01}
  \dot x= X(x), \qquad x\in M
\end{equation}
having first integrals
\begin{equation}\label{eq02}
  K_1,K_2,\ldots,K_n: M \to \bR.
\end{equation}

In connection with the analysis of the trajectories of system \eqref{eq01}, S.\,Smale [1] formulates the problem of investigation of the topological types of integral manifolds
\begin{equation}\label{eq03}
  J_k =\{x\in M: K_i(x)=k_i, i=1,2,\ldots,n\}.
\end{equation}
Here $k=(k_1,k_2,\ldots,k_n)\in \bR$ stands for the set of arbitrary integral constants. This problem can be solved by classifying the projections of $J_k$ to some manifold $N$ such that $\dim N <\dim M$. In mechanics, this manifold arises in natural way. It is the space of variables depending only on the object configuration. The investigation of the projections of phase trajectories to the space of configuration variables is also of special interest because these projections are straightforwardly connected with the motion of the real mechanical system.

Let $M$ be the total space of a locally trivial vector bundle $\pi: M \to N$ with a fiber $F$. For $z\in N$, we denote $F_z=F^{-1}(z)$.

\vskip2mm
\textbf{Definition}. \textit{The set $u_k=\pi(J_k)$ is called the} \textbf{region of possible motion} (\textit{RPM}) \textit{on $N$}. \textit{Any vector $w\in F_z\cap J_k$ is called an} \textbf{admissible velocity} \textit{at the point $z\in u_k$.}
\vskip2mm

Denote
\begin{equation}\label{eq04}
  \pi_k=\pi|_{J_k}:J_k \to N.
\end{equation}

Two regions of possible motion $u_k,u_{k'}$ $(k,k'\in \bR^n)$ are considered equivalent if there exists a diffeomorphism $\Gamma:J_k \to J_{k'}$ such that $\pi_{k'}\circ\Gamma=\pi_k$.
\vskip2mm
\textbf{Definition}. \textit{For a given RPM $u_k$ the visible contour} \textit{$\partial_k$ of the manifold $J_k$ under the projection $\pi_k$ is called the} \textbf{generalized boundary} \textit{of $u_k$}.
\vskip2mm

In other words, the visible contour [2,3] can be defined as the set $\partial _k$ of those points of $N$ over which the map \eqref{eq04} is not locally trivial. Obviously, if $u_k$ and $u_{k'}$ are equivalent, then there exists a diffeomorphism $\gamma:u_k \to u_{k'}$ such that $\gamma(\partial_ {k})=\partial_{k'}$.

If $\dim J_k \geqslant \dim N$, then $\partial_k$ divides $N$ into open connected components, inside which the structure of the sets of the admissible velocities does not change. Pointing out these sets at some point of each component and at the points of $\partial_k$, we present the integral manifold $J_k$ in the form of some bundle over $u_k$.

Let $x\in M, z=\pi(x)$. We denote by $V(x):F\to \bR^n$ the restriction to the subspace $T_x F_z\subset T_x M$ ($T_x F_z \cong F$) of the operator $T_x(K_1{\times}K_2{\times}\ldots {\times}K_n):T_x M \to T_k \bR^n \cong \bR^n$.

\vskip2mm
\textbf{Theorem}. \textit{The image of a point $x\in J_k$ belongs to the generalized boundary of the corresponding RPM if and only if}
\begin{equation}\label{eq05}
  \mathop{\rm rank}\nolimits V(x)<n.
\end{equation}
\vskip2mm

In particular, the generalized boundary always contains the image of those points of $J_k$, at which the integrals \eqref{eq02} are dependent.

The suggested method of reconstructing the integral manifolds makes it possible to describe the surfaces \eqref{eq03} also in those cases when they are not smooth [4,5]. This is important, first, because there are no general theorems on the structure of critical levels of smooth mappings, and, second, because the special classes of motions always belong to the critical integral surfaces~\eqref{eq03}.

\vskip3mm

\def\oq{\omega}
\def\lm{\lambda}
\def\sq{\sigma}

\begin{table}
\centering
{\small
\begin{tabular}{cc}
\includegraphics[width=0.3\textwidth]{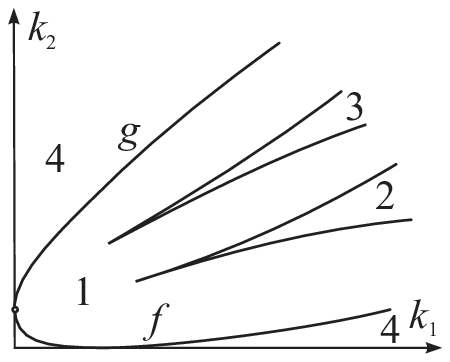} & \includegraphics[width=0.5\textwidth]{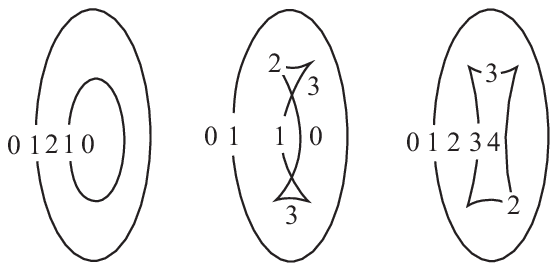} \\
{Figure}~1 & {Figure}~2
\end{tabular}
}
\end{table}

E x a m p l e. Consider the problem of the motion of a gyrostat without external forces. In the moving frame, this problem is described by the equations
\begin{equation}\label{eq06}
  A \dot \oq +\oq{\times}(A\oq+\lm)=0, \qquad \dot \nu =\nu{\times}\oq.
\end{equation}

Usually, the following relation is accepted
\begin{equation}\label{eq07}
  \nu \cdot \nu =1.
\end{equation}
It defines the so-called Poisson sphere. The phase space of system \eqref{eq06} then is $M=S^2{\times}\bR^3$. The system on $M$ has three first integrals [6]
\begin{equation*}
  K_1 = (A \oq+\lm)\cdot(A \oq+\lm), \quad K_2=A\oq\cdot\oq, \quad K_3=(A \oq+\lm)\cdot \nu.
\end{equation*}
Their common levels $J_k$ ($k\in \bR^3$) generally are two-dimensional surfaces.

Let us suppose that the gyrostatic moment $\lm$ does not belong to any of the the principal planes of the tensor $A$. The set $\Sigma$ of the points $k\in \bR^3$ corresponding to the changes in the structure of $J_k$ can be described as follows. On the plane $(k_1,k_2)$, we introduce the curve (see Fig.~1)
\begin{equation}\label{eq08}
  \left\{
\begin{array}{ll}
\ds  k_1= \frac{a_1^2 \lm_1^2}{(\sq-a_1)^2}+\frac{a_2^2 \lm_2^2}{(\sq-a_2)^2}+\frac{a_3^2 \lm_3^2}{(\sq-a_3)^2}, & -\infty \leqslant \sq \leqslant\infty,\\
\ds k_2=\sq^2\left[\frac{a_1 \lm_1^2}{(\sq-a_1)^2}+\frac{a_2 \lm_2^2}{(\sq-a_2)^2}+\frac{a_3 \lm_3^2}{(\sq-a_3)^2} \right], &  \sq\ne a_1,a_2,a_3.
\end{array}
  \right.
\end{equation}
Here $a_1,a_2<a_3$ are the diagonal elements of the tensor $A^{-1}$ in the principle axes. Denote by $k_2=f(k_1)$ and $k_2=f(k_1)$ the one-valued branches of the curve \eqref{eq08} corresponding to the segments $-\infty \leqslant \sq < a_1$ and $a_3< \sq \leqslant\infty$ respectively. Let $C_1$ be the cylinder in $\bR^3$ with the directrix \eqref{eq08} and generators parallel to $Ok_3$. Denote by $C_2$ the parabolic cylinder $k_1=k_3^2$. Then
\begin{equation*}
  \Sigma = \{k\in C_1: k_1 \geqslant k_3^2\} \cup \{k\in C_2: f(k_1)\leqslant k_2 \leqslant g(k_1)\}.
\end{equation*}
This set divides $\bR^3$ into four regions 1\,--\,4; in each region the type of the manifold $J_k$ is constant. Obviously, $J_k=\varnothing$ in region 4.

Let us take the sphere \eqref{eq07} for the manifold $N$. The condition \eqref{eq05} takes the form
\begin{equation}\label{eq09}
  \left[\oq{\times}(A\oq+\lm)\right]\cdot \nu =0.
\end{equation}
Together with the equations
\begin{equation}\label{eq10}
  K_1(\oq)=k_1, \quad K_2(\oq)=k_2,\quad K_3(\oq,\nu)=k_3
\end{equation}
equality \eqref{eq09} defines a curve in $M$; the image of this curve on the sphere \eqref{eq07} is the generalized boundary of RPM:
\begin{equation}\label{eq11}
\ds  \nu =\left[k_3 \pm (k_2+\oq\cdot\lm)Q(\oq)\right]\frac{(A\oq+\lm)}{k_1}\pm Q(\oq)\oq.
\end{equation}
Here
\begin{equation*}
\ds   Q(\oq)= \sqrt{\frac{k_1-k_3^2}{k_1\oq^2-(k_2+\oq\cdot\lm)^2}}\,.
\end{equation*}

From the first two equations \eqref{eq10} we can express the projections of the angular velocity in one auxiliary variable $\tau$ (see [7]) as $\oq_i=\oq_i(\tau; k_1,k_2)$ ($i=1,2,3$). Substituting these values in the right-hand part of \eqref{eq11} we obtain the explicit parametric equations of the generalized boundary. Investigating these equations, we come to the main types of the connected components of an RPM shown in Fig.~2. The numbers on the figure show the number of the admissible velocities. Obviously, we have the images of the two-dimensional torus $T^2$. All of these images are found in any of the regions 1\,--\,3 in $\bR^3\backslash \Sigma$. In region 1 the RPM consists of one component of the shown type, in regions 2,3 we have two such components. Thus, $J_k=T^2$ in region 1 and $J_k=2T^2$ in regions 2 and 3. The trajectories on the tori are quasi-periodic [2], therefore the character of the motions on the Poisson sphere is also clear. At the points of $\Sigma$ the integral surfaces have one of the following types: the circle, the direct product of the eight-type curve with the circle or with the circle having an angular point, the disjoint union of the circle and the two-dimensional torus bearing quasi-periodic motions.

\vskip5mm

\flushright{Translated by M.P.Kh.}

Address: \textit{Russia, 400131, Volgograd, Gagarin Street, 8, Volgofrad Branch of RANEPA}

E-mail: \textit{mharlamov@vags.ru}


\begin{thebibliography}{9}

\bibitem{Sm1970}
S. Smale, Topology and mechanics, {\it Inventiones Math.} {\bf 10}, 4 (1970), pp.~305--331.

\bibitem{Ar1989}
V.I. Arnold, Mathematical Methods of Classical Mechanics (Springer, 1989).

\bibitem{Pham}
F. Pham, Introduction a l'\'{E}tude Topologique des Singularit\'{e}s de Landau (Paris, Gauthier-Villars \'{E}diteur, 1967).

\bibitem{Kh1979}
M.P. Kharlamov, Phase topology of one integrable case of the rigid body motion, {\it Mekh. Tverd. Tela}, No.~11 (1979), pp.~50--64. (In Russian)

\bibitem{Kh1981}
M.P. Kharlamov, Phase topology of one problem of motion of a gyroscope, {\it Mekh. Tverd. Tela}, No.~13 (1981), pp.~14--23. (In Russian)

\bibitem{Zh1949}
N.E. Zhukovsky, On the motion of a rigid body with holes filled with a homogeneous fluid, in {\it Collected Works}, vol. 1 (1949), pp.~31--152. (In Russian)

\bibitem{Witt}
J. Wittenburg, Dynamics of Systems of Rigid Bodies (B.G. Teubner Stuttgart, 1977).

\end{thebibliography}
\end{document}